\documentclass[aps,pra,twocolumn,amsmath,amssymb]{revtex4}

\usepackage{graphicx}
\usepackage{amsmath}

\usepackage{breqn}

\usepackage{braket}

\usepackage{bbold}
\usepackage{color}

\usepackage[spanish, activeacute, english]{babel}
\spanishdecimal{.}
\usepackage[utf8]{inputenc}

\usepackage[colorlinks = true,
            linkcolor = blue,
            urlcolor  = blue,
            citecolor = blue,
            anchorcolor = teal]{hyperref}

\usepackage[small]{caption} 
\usepackage{float}
\usepackage{xcolor}

\usepackage{tocloft}

\begin{document}

\title{Synchronization in two-level quantum systems}

\author{ Álvaro Parra-López }

\author{Joakim Bergli }
\affiliation{Department of Physics$,$ University of Oslo$,$ NO-0316 Oslo$,$ Norway}

\selectlanguage{english}

\date{\today}

\begin{abstract}
 Recently \cite{paper, paper2, paper3}, it was shown that dissipative quantum systems with three or more levels are able to synchronize to an external signal, but it was stated that it is
not possible for two-level systems as they lack a stable limit
cycle in the unperturbed dynamics. At the same time, several papers  \cite{TPaper1, TPaper2, ExtraPaper3, ExtraPaper4}, demonstrate, under a different definition of what is synchronization, that the latter is possible in qubits, although in  different models which also include other elements. We show how a quantum two-level system can be understood as containing a valid limit cycle as the starting point of synchronization, and that it can synchronize
its dynamics to an external weak signal. This is demonstrated by analytically solving
the Lindblad equation of a two-level system coupled to an environment, determining the steady state. This is a mixed state with contributions from many pure states, each of
which provides a valid limit cycle. We show that this is
sufficient to phase-lock the dynamics to a weak
external signal, hence clarifying synchronization in two-level systems. We use the Husimi Q representation to analyze the synchronization region, defining a synchronization measure which
characterizes the strength of the phase-locking. Also, we study the stability of the limit cycle and its deformation with the strength of the signal in terms of the components of the Bloch vector of the system. Finally, we generalize the model of the three-level system from \cite{paper} in order to illustrate how the stationary fixed point of that model can be changed into a limit cycle similar to the one that we describe for the two-level system.
\end{abstract}

\maketitle

\section{Introduction}

The phenomenon of synchronization occurs in many different situations and has been extensively studied for many years. If an autonomous oscillating system is coupled to another such system or to an external driving force it can synchronize its frequency and phase to the external system. Examples are coupled pendulums, circadian rhythms  in living systems or synchronization of fireflies flashing. Common to these systems are the fact that they need to have a stable limit cycle, which means they must be dissipative, so that they can return to the stable cycle after a perturbation, and contain an energy source, so that they can sustain oscillations indefinitely in the presence of dissipation \cite{book}.

One of the well studied examples of classical synchronization is the van der Pol oscillator model 
\cite{book, vanderc}. Some years ago, the van der Pol model was reformulated in terms of a quantum system \cite{vander, vander2}, and it was shown  that when the system is far from the ground state, synchronization in quantum systems is analogous to classical synchronization of the same system in the presence of noise \cite{book}. When we are close to the ground state, this correspondence is changed because the discreteness of the energy levels becomes important. It is therefore interesting to study synchronization in quantum systems with a small number of energy levels. 

The natural idea is to synchronize a two-level system (TLS) either with another two-level system, as it was done in Refs.~\cite{TPaper1, TPaper2}, or with an external signal.  The latter was discussed in Ref.~\cite{paper}, with the conclusion that it is not possible to have a stable limit cycle in the dynamics of a dissipative TLS, and therefore synchronization can not occur. However, in \cite{ExtraPaper3, ExtraPaper4} it is claimed that synchronization between a qubit and a driven external signal is possible, although there is no illustration about what is the limit cycle of the system or how it does occur. A brief discussion of the synchronization of a single qubit to an external signal is also present in Ref.~\cite{2017arXiv170504614E}. We show how one can understand the appearance of a valid limit cycle, which is an essential starting point for synchronization,  if one aims at relating the quantum version of this phenomena to its classical counterpart. In this context, the system is not completely phase locked, therefore if we accept that the quantum system is similar to a classical system with noise, as is also the case for the van der Pol oscillator~\cite{vander, vander2} (and all quantum systems in general \cite{ExtraPaper1, ExtraPaper2, ExtraPaper5}), a TLS is in fact capable of synchronization, and the following considerations allow us to understand why. 
 
Our system is in contact with an environment so that it is able to gain and emit energy, hence creating the dissipating frame synchronization requires.
Solving the Lindblad equation for the system in the absence of any external signal one finds that the stationary solutions are mixed states that are constant in time lying on the rotation axis of the Bloch sphere, which we will take to be the $z$-axis. As was stated in Ref.~\cite{paper}, this is not a valid limit cycle and it seems that it can not form the starting point for synchronization. Nevertheless, as these states are mixed, we must understand that the system is in a probability mixture of some pure states. While the ensemble of pure states which generates a given mixed state is not unique, we can choose them to be on that circle on the surface of the Bloch sphere which is in the plane normal to the $z$-axis and which has the given mixed state at the center, see Fig.~\ref{2DBS}.  
Each of these states is then rotating and provides a limit cycle, while the mixed state of the ensemble is stationary. This argument does not apply to the three-level system discussed in Ref. \cite{paper}, since in that case the steady state in the absence of an external signal is an eigenstate of the Hamiltonian, and only when we  introduce an external signal do we  obtain a cycle by driving the system away from this state. We will illustrate at the end how to modify this model so that the stationary state in the absence of a signal is a limit cycle in the same sense as for the TLS we consider here. 
In the following we will show that a two-level system indeed displays all the signatures of synchronization as was already demonstrated for a three-level system (spin-1)~\cite{paper}, but also that it is possible to understand this synchronization as arising from the limit cycles provided by the mixed steady state in the unperturbed dynamics.

\section{Model and limit cycle}

We consider a two-level system characterized by the Hamiltonian

\begin{equation}
\hat{H}_0 = \frac{\hbar}{2} \omega _0 \mathbf{n} \cdot \boldsymbol{\hat{\sigma}}
\end{equation}
where $\boldsymbol{\hat{\sigma}}$ is a vector containing the three Pauli matrices. 
The density operator for this system can be written as

\begin{equation}
\hat{\rho'} = \frac{1}{2}(\mathbb{1} + \mathbf{m}' \cdot \boldsymbol{\hat{\sigma}})
\end{equation}
where $\mathbf{m}'$ is the Bloch vector \cite{bellac}.
Whenever $\mathbf{m}'$ is not pointing in the same direction as $\mathbf{n}$,
the Bloch vector precesses around the \textit{n}-axis with frequency $\omega _0$, which we call the \textit{natural frequency} of the system.
However, although this may indicate the existence of a limit cycle, states should be stable under perturbations. Thus, we must consider a model in which the system is capable of gaining and losing energy.
If we choose $\mathbf{n}$ such that it points towards the $z$-direction in the Bloch sphere, we can write the Hamiltonian as

\begin{equation}
\hat{H}_0 = \frac{\hbar}{2}\omega _0 \hat{\sigma} _z
\end{equation} 

We transform to a frame rotating with the natural frequency $\omega_0$, defining the density matrix in the rotating frame as $\hat{\rho} = \hat{T}_{\omega _0} \hat{\rho'}\hat{T}_{\omega _0}^\dagger $
where 
\begin{equation}
\hat{T}_{\omega _0} = e^{i\frac{\omega _0}{2}\hat{\sigma}_z t},
\end{equation}
and denote the corresponding Bloch vector $\mathbf{m}$.
In this frame, the Lindblad equation including gain and damping  is \cite{bellac}

\begin{dmath}
\frac{d\hat{\rho}}{dt} = \frac{\Gamma _g}{2} \mathcal{D}[\hat{\sigma}_+]\hat{\rho} + \frac{\Gamma _d}{2} \mathcal{D}[\hat{\sigma}_-]\hat{\rho} 
\end{dmath}  
where $\Gamma _g$ and $\Gamma _d$ are the gain and damping rates, $\mathcal{D}[\hat{\mathcal{O}}]\hat{\rho} = \hat{\mathcal{O}}\hat{\rho}\hat{\mathcal{O}}^{\dagger} - \frac{1}{2}\{\hat{\mathcal{O}}^{\dagger}\hat{\mathcal{O}}, \hat{\rho}\}$ is the Lindblad superoperator and $\hat{\sigma}_+$ and $\hat{\sigma}_-$ are the ladder operators for the system, $\hat{\sigma}_{\pm} =\frac{1}{2}(\hat{\sigma}_x \pm i\hat{\sigma}_y)$.  
This equation is equivalent to the one studied in Ref.~\cite{paper}.

In terms of the Bloch vector components, we find the following equations:

\begin{dgroup}[style = {\small}]
\begin{dmath}
\dot{m}_x = -\frac{1}{4}(\Gamma _d + \Gamma _g)m_x
\end{dmath}

\begin{dmath}
\dot{m}_y = -\frac{1}{4}(\Gamma _d + \Gamma _g)m_y
\end{dmath}

\begin{dmath}
\dot{m}_z = \frac{1}{2}[\Gamma_g(1-m_z) - \Gamma_d(1 + m_z)]
\end{dmath}

\end{dgroup}
As we are working in a frame rotating with the natural frequency of the system, a point that precesses in the non-rotating frame should be now a fixed point. Thus, we look for stationary solutions, i.e., $\hat{\dot{\rho}} = 0$, $\mathbf{\dot{m}} = 0$. The solution is then

\begin{equation}
m_x = 0 ;\quad m_y = 0 ; \quad m_z = \frac{\Gamma _g -\Gamma _d}{\Gamma_g + \Gamma _d}
\end{equation}

For the ground state $\ket{\downarrow}$ and the excited state $\ket{\uparrow}$ which correspond to the Bloch vectors $\mathbf{m}_g = (0, 0, -1)$ and $\mathbf{m}_e = (0, 0, 1)$, respectively, we do not expect a limit cycle, as the state is a fixed point. However, even if the solution for any of the other cases is lying on the $z$-axis, we must remember that they are  mixed states. This means that our solution is a mixture of pure states, each of them weighted with a certain probability. It is not a superposition and our system is for sure in any of those pure states, but \textit{only in one of them at the same time}. 

\begin{figure}[H]
\centering
\includegraphics[width=8cm]{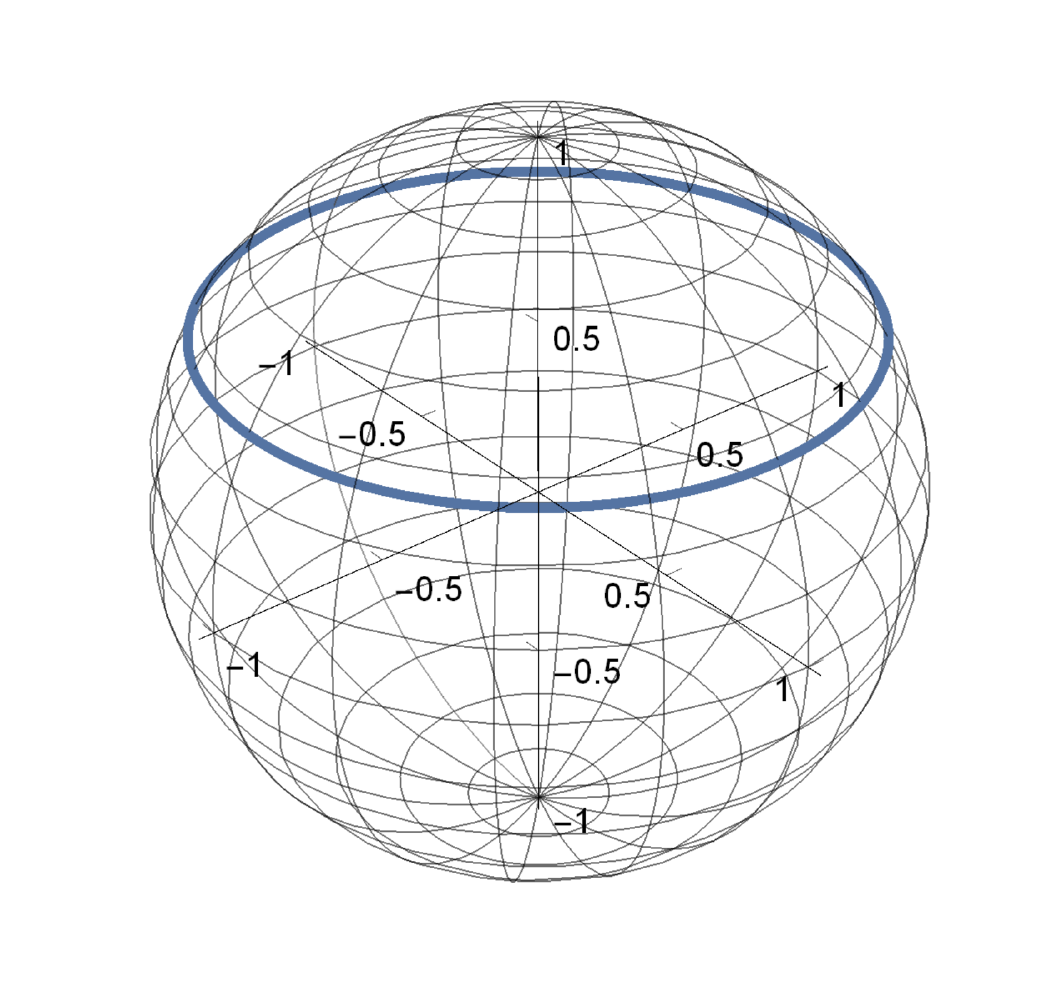}
	\caption{We can realize the mixed steady state as a probabilistic ensemble of the states located on the surface of the Bloch sphere (state space) in the plane normal to the $z$-axis (blue circle). Each of these states move along the same circle, with a periodic motion that will correspond to a limit cycle in the phase space. In this example, we have chosen $\frac{\Gamma_g}{\Gamma_d} = 3$.}
	\label{2DBS}
\end{figure}

Thus, the limit cycle is provided given that each of those possible pure states that make up our mixed state would precess around the $z$-axis once we move back to the non-rotating frame. For example, a mixture of states lying on a circle on the surface of the Bloch sphere, in a plane normal to the $z$-axis and with the steady state in the centre, as it is illustrated in Fig. \ref{2DBS}. However, the stationary state can always be realized as a mixture of the pure states $\ket{\downarrow}$ and $\ket{\uparrow}$, which are not precessing, and this means that the above argument is not fully convincing. Having this picture in mind, we will now demonstrate that a TLS indeed allows synchronization in the presence of an external signal.

\section{Synchronization of the TLS}
In order to synchronize our system with an external signal, we use a classical drive \cite{paper} of frequency $\omega$ and strength $\epsilon$. In the rotating-wave approximation \cite{bellac}, it is given by the Hamiltonian

\begin{dmath}
\hat{H}_{signal} = i \hbar \frac{\epsilon}{4}(e^{i\omega t} \hat{\sigma}_- - e^{-i \omega t}\hat{\sigma}_+)
\end{dmath}

If we want to move to a frame rotating with the frequency of the signal, the transformation operator we must apply to our Lindblad equation is

\begin{equation}
\hat{T}_{\omega} = e^{i\frac{\omega}{2}\hat{\sigma}_z t} 
\end{equation}
giving

\begin{dmath}[style = {\small}]\label{LindbladTransf2LS}
\frac{d\hat{\rho}}{dt} = -\frac{i}{2}[\Delta \hat{\sigma}_z + \epsilon \hat{\sigma}_y, \hat{\rho}] + \frac{\Gamma _g}{2} \mathcal{D}[\hat{\sigma}_+]\hat{\rho} + \frac{\Gamma _d}{2} \mathcal{D}[\hat{\sigma}_-]\hat{\rho} 
\end{dmath}
where $\Delta = \omega _0 - \omega$.

Again, we obtain the evolution equations for the Bloch vector components,

\begin{dgroup}[style = {\small}]
\begin{dmath}
\dot{m}_x = -\frac{1}{4}(\Gamma _d + \Gamma _g)m_x - \Delta m_y + \epsilon m_z
\end{dmath}

\begin{dmath}
\dot{m}_y = \Delta m_x -\frac{1}{4}(\Gamma _d + \Gamma _g)m_y
\end{dmath}

\begin{dmath}
\dot{m}_z = \frac{1}{2}[\Gamma_g(1-m_z) - \Gamma_d(1 + m_z) - 2 \epsilon m_x]
\end{dmath}
\end{dgroup}
with the stationary solution 

\begin{dgroup}[style = {\small}]\label{stationarySolution}
\begin{dmath}
m_x = \frac{4\epsilon (\Gamma _g - \Gamma _d)}{(\Gamma _d + \Gamma_ g)^2 + 8(\epsilon ^2 + 2\Delta ^2)}
\end{dmath}

\begin{dmath}
m_y = \frac{16\epsilon \Delta(\Gamma _g - \Gamma _d)}{(\Gamma_d + \Gamma _g)[(\Gamma _d + \Gamma _g)^2 + 8(\epsilon ^2 + 2\Delta ^2)]}
\end{dmath}

\begin{dmath}
m_z = \frac{(\Gamma _d - \Gamma _g)[(\Gamma _d + \Gamma _g)^2 + 16 \Delta ^2]}{(\Gamma _d + \Gamma _g)[(\Gamma _d + \Gamma _g)^2 + 8(\epsilon ^2 + 2\Delta ^2)]}.
\end{dmath}
\end{dgroup}

In order to obtain the state operator in the non-rotating frame, we transform back with $\hat{T}_{\omega} $, and thus the state operators are related by $\hat{\rho}^' = \hat{T}_{\omega} ^{\dagger}\hat{\rho}\hat{T}_{\omega}$. Therefore, the Bloch vector in the non-rotating frame will be given by

\begin{dgroup}\label{nonrotvec}

\begin{dmath}
m'_x = m_x \cos{\omega t} - m_y \sin{\omega t}
\end{dmath}

\begin{dmath}
m'_y = m_x \sin{\omega t} + m_y \cos{\omega t}
\end{dmath}

\begin{dmath}
m'_z = m_z
\end{dmath}

\end{dgroup}
When transforming back to the non-rotating frame, $m'_x$ and $m'_y$ will vary in time with the frequency of the signal, 
which means that the system phase-locks to the external force.

Eq.~\eqref{stationarySolution} give steady states with non-zero transverse Bloch vector components $m_x$ and $m_y$, provided the damping and gain rates are not equal and the strength of the signal is different than zero.
Thus, Eq.~\eqref{nonrotvec} will give a precessing vector with frequency $\omega$ in the non-rotating reference frame. Also, it is not difficult to show that it does not matter which is the initial state: after some transient, the motion will be the one described by the steady solution.

When both $m_x$ and $m_y$ are zero, the steady state is lying on the $z$-axis, and will still be a fixed point in the non-rotating frame ($m'_z = 0$ if $\Gamma _g = \Gamma _d$). Therefore, there is no synchronization in this case. 

In order to visualize the behavior of the system, we follow Ref.~\cite{paper} and use the 
Husimi Q representation adapted to spin systems \cite{cohe}. This is a quasi-probability distribution that allows us to represent the phase space of the two-level system and is defined by 

\begin{equation}
Q(\theta, \phi) = \frac{1}{2 \pi} \bra{\theta, \phi}\hat{\rho}\ket{\theta, \phi}.
\end{equation}
Here $\ket{\theta, \phi}$ are spin-coherent states, which in the case of a TLS are the eigenstates of the spin operator $\sigma_{\mathbf n} = \mathbf{n} \cdot \boldsymbol{\hat{\sigma}}$ along the axis given by the unit vector $\mathbf{n}$ which has polar coordinates $\theta$ and $\phi$.
These are 
 nothing but the pure states at the corresponding point on the Bloch sphere in terms of the angles $\theta$ and $\phi$. Therefore, what the Q representation is telling us is how every pure state (corresponding to a pair of angles in the Bloch sphere) that contributes to the state operator $\hat{\rho}$ is weigthed:

\begin{equation}
\bra{\theta, \phi}\hat{\rho}\ket{\theta, \phi} = \sum _n P_n |\braket{\theta, \phi|\psi_n}|^2
\end{equation}
where $\hat{\rho} = \sum _n P_n \ket{\psi _n}\bra{\psi _n}$.

Given the solution, Eq.~\eqref{stationarySolution}, it is easy to find that the Q-function of the steady states, as a function of the components of the Bloch vector, is

\begin{dmath}
Q(\theta, \phi) = \frac{1}{4 \pi}[1 + m_x \cos{\phi}\sin{\theta} + m_y \sin{\phi}\sin{\theta} + m_z \cos{\theta}]
\end{dmath}

\begin{figure}[H]
	\includegraphics[width=7cm]{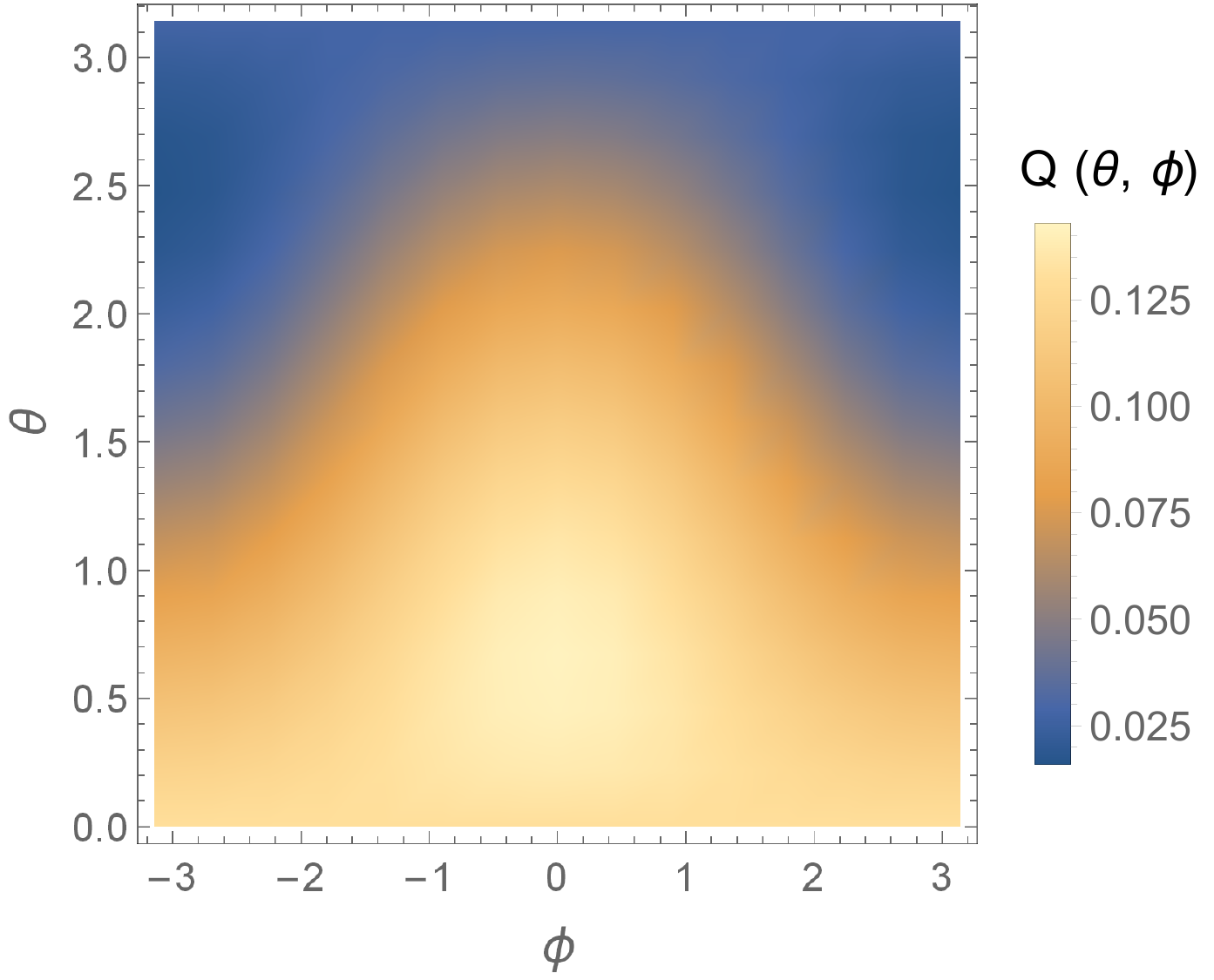}
	\caption{Q-function of the steady state for $\epsilon = 2 \cdot min\{\Gamma _g, \Gamma _d\}$, $\Delta = 0$, $\frac{\Gamma _g}{\Gamma _ d} = 10$. The distribution is peaked around $\theta = 0$ because it is the gain rate that dominates, and around $\phi = 0$ because we are synchronizing to a resonant signal (there is no detuning). Complete phase-locking does not occur and the Q-function is non-zero everywhere. Note that the units of $\epsilon$ and $\Delta$ will depend on those chosen for the rates.}
	\label{QFe}
\end{figure}

\begin{figure}[H]
	\includegraphics[width=7cm]{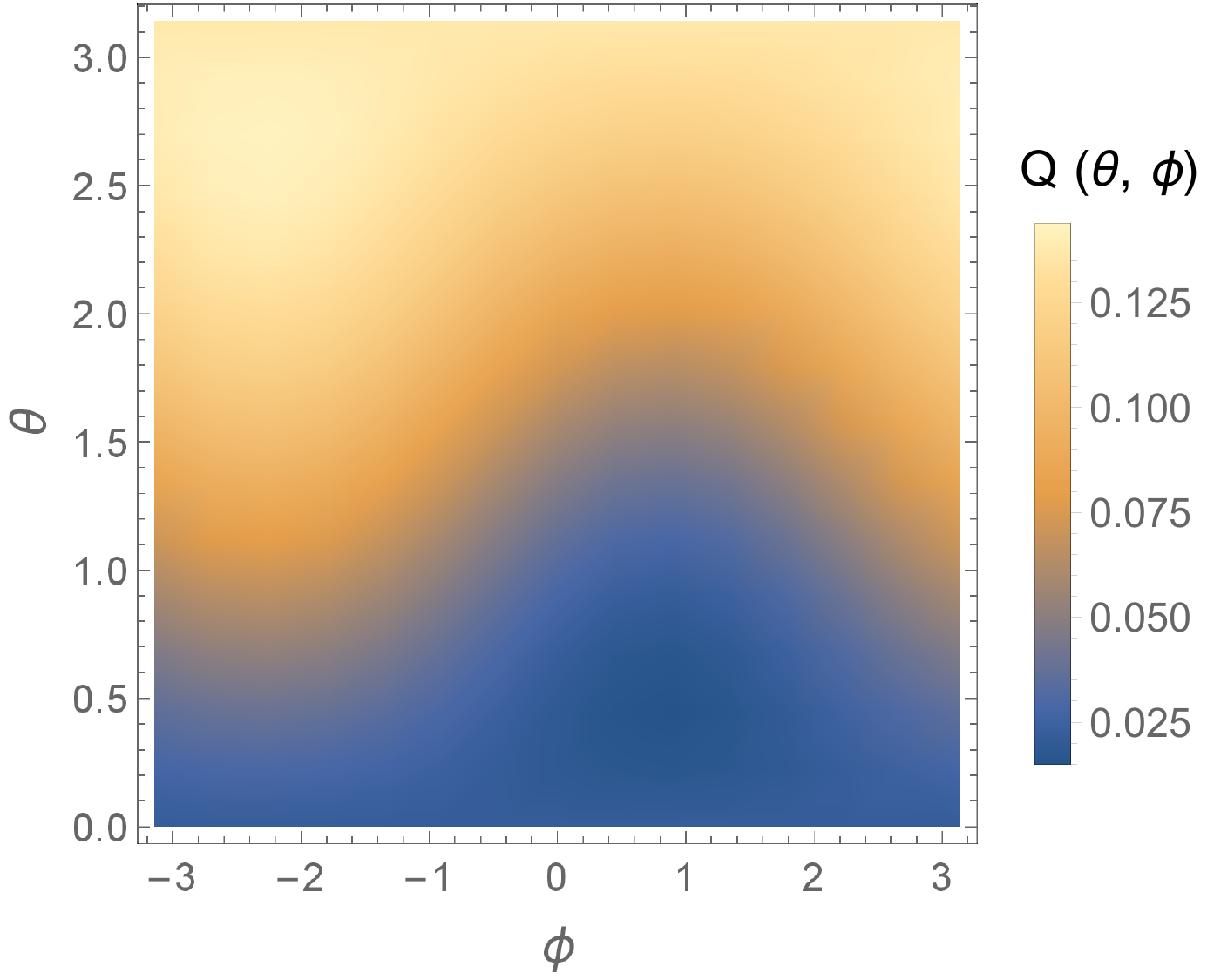}
	\caption{Q-function of the steady state for $\epsilon = 2 \cdot min\{\Gamma _g, \Gamma _d\}$, $\Delta = 3 \cdot min\{\Gamma _g, \Gamma _d\}$, $\frac{\Gamma _g}{\Gamma _ d} = 0.1$. Since $\Gamma_d>\Gamma_g$  we find the higher values of the function around $\theta = \pi$. Because of the detuning, the distribution is displaced along the $\phi$-axis, moving away from $\phi = \pi$, which is where it would be located if $\Delta = 0$.}
	\label{QFd}
\end{figure}

Fig. \ref{QFe} shows the Q-function for the case where there is no detuning and the gain rate is larger than the damping rate, $\Gamma_g>\Gamma_d$. As expected, the states that contribute the most to the mixture of the steady state are those corresponding to $\theta = 0$. Also, the distribution is located around $\phi = 0$. The system is phase-locked in the sense that the state is made up mostly by contributions from a specific $\phi$ region.

On the other hand, in Fig. \ref{QFd}, it is the damping rate that dominates. Hence, we expect higher values of the Q-function at $\theta$ values close to $\pi$. When this is the case, for no detuning, the distribution is situated at $\phi = \pi$. However, in Fig. \ref{QFd} the detuning shifts the phase towards $\phi = \frac{\pi}{2}$. In terms of the Bloch vector, the detuning makes $m_y$ non-zero, therefore the projection in the $xy$-plane is not a vector lying on the $x$-axis, which is the case for Fig. \ref{QFe}, where there is no detuning. 

As a next step, we attempt to measure how strong is synchronization defining a \textit{synchronization measure}.

\section{Synchronization region}

Even if we are able to observe phase-locking when plotting the Q-function, we would like to characterize its strength. There is a tool that allows us to do it, and following the work done in \cite{paper}, we define a \textit{synchronization measure},

\begin{equation}
S(\phi) = \int _0 ^{\pi} d\theta \sin{\theta}Q(\theta, \phi) - \frac{1}{2\pi}
\end{equation}
This is identically zero when there is no synchronization, i.e., when only $m _z$ is non-zero. Explicitly performing the integral over $\theta$ we find that

\begin{equation}
S(\phi) = \frac{1}{8}( m_x \cos{\phi} + m_y \sin{\phi})
\end{equation}
Thus, $S(\phi)$ is going to be greater as the steady state is farther from the $z$-axis.

\begin{figure}[H]
	\includegraphics[width=8cm]{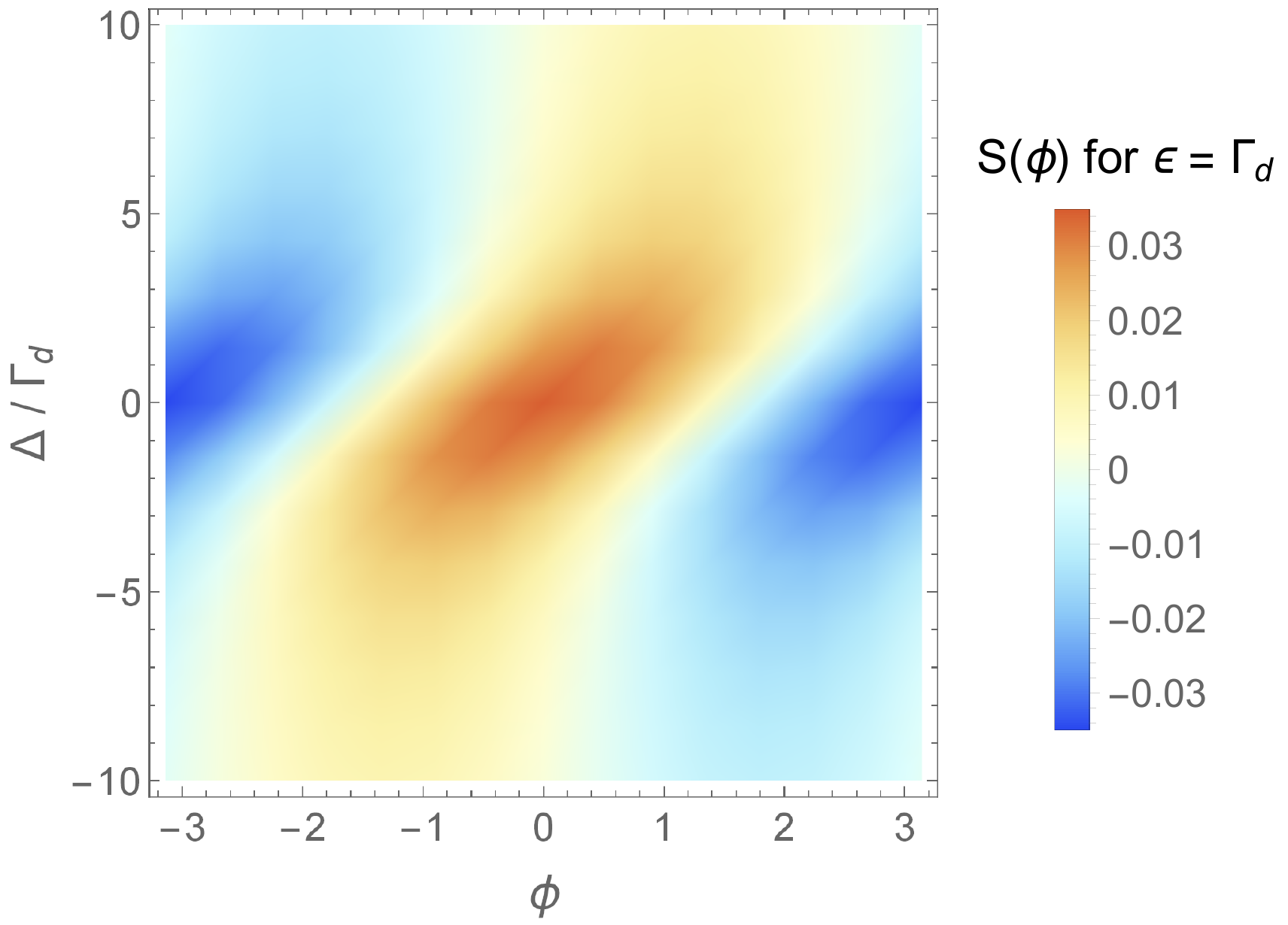}
	\caption{$S(\phi)$ for $\epsilon = \Gamma _d$ for different $\Delta$, $\frac{\Gamma _d}{\Gamma _g} = 0.1$. As expected, the phase-locking is stronger when there is no detuning. When $\Delta$ is positive or negative, the maximal value of $S(\phi)$ is shifted  towards $\phi = \pi$ or $\phi = -\pi$, respectively.}
	\label{ATa}
\end{figure}

\begin{figure}[H]
	\includegraphics[width=8cm]{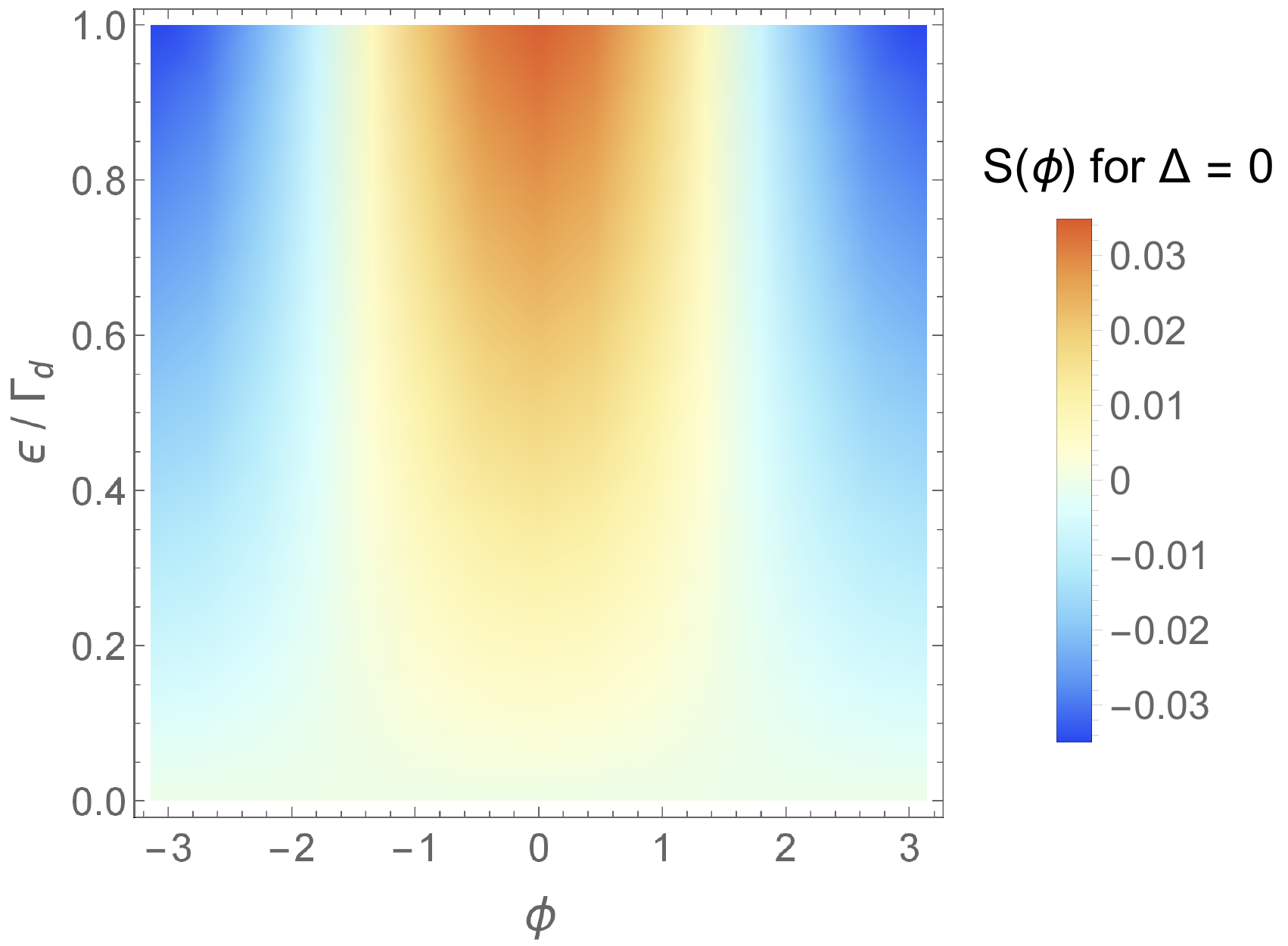}
	\caption{$S(\phi)$ for $\Delta = 0$ for different $\epsilon$. We study how the strength of the signal modifies the strength of the phase-locking. We observe that the greater the strength of the signal, the strongest phase-locking occurs. It is because there is no detuning that the highest values locate at $\phi = 0$ (in-phase synchronization)  as we could deduce from Fig. \ref{ATa}.}
	\label{ATb}
\end{figure}

\begin{figure}[H]
	\includegraphics[width=8cm]{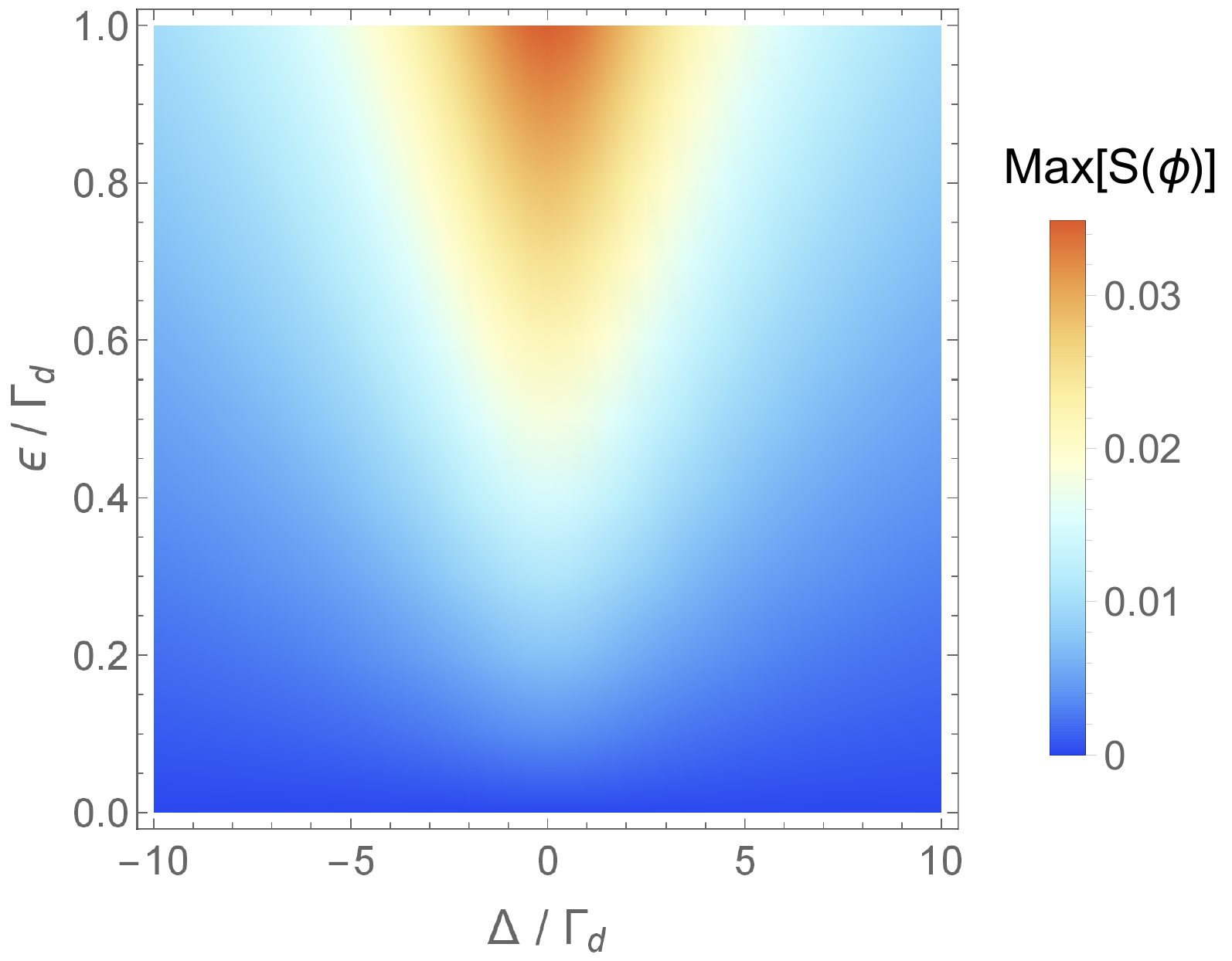}
	\caption{Arnold Tongue of the system. We plot the maximum value of $S(\phi)$ as function of the strength $\epsilon$ and the detuning $\Delta$, with $\frac{\Gamma _d}{\Gamma _g} = 0.1$. This plot resembles the one in Ref. \cite{paper}. Unless $\epsilon = 0$ or $\Gamma_g = \Gamma_d$, \eqref{stationarySolution} will always give non-zero $m_x$ and $m_y$, and therefore $Max[S(\phi)]$ will be non-zero, even if it is so small that the synchronization is negligible.}
	\label{ATc}
\end{figure}

Figures \ref{ATa}-\ref{ATc} are useful for understanding the synchronization dynamics. In Fig. \ref{ATa} we can observe how the detuning drives the phase of the system towards positive or negative $\phi$ if the detuning ($\Delta = \omega _0 - \omega$) is positive or negative. The larger the absolute value of $\Delta$ is, the weaker the synchronization will be. Note that for this figure we have $\frac{\Gamma _d}{\Gamma _g} <1$ and we observe in-phase synchronization. If instead we use $\frac{\Gamma _d}{\Gamma _g} >1$ we would observe anti-phase synchronization, which means that the maximal value of $S(\phi)$ will be at $\phi=\pi$ for $\Delta=0$.

On the other hand, Fig. \ref{ATb} takes into account the effect of growing signal strength on the synchronization measure. It is clear that the synchronization is stronger for a greater $\epsilon$ (both in- and anti-phase synchronization), but we must keep in mind that a very strong signal would take us out of the synchronization regime.

Finally, the \textit{Arnold Tongue} that is characteristic of every synchronized system is displayed in Fig. \ref{ATc}.  The shape is that of the \textit{tongue} for a spin-1 system (Fig. 3 from \cite{paper}), and it is worth to mention that $S(\phi)$ is different from zero everywhere except for the case when $\epsilon = 0$.   
The  state will always precess with the frequency of the signal but depending on the size of $m_x$ and $m_y$, the synchronization measure will be greater or smaller, telling us 
how strong  the phase-locking is. 

We have seen that the results for the two-level system are equivalent to those obtained in Ref. \cite{paper} for the three-level system. Moreover, we have explained the appearance of the limit cycle for the present case.
Nevertheless, in order to stay in the synchronization regime, the strength of the external signal should not be too large, and we will analyse how the limit cycle is distorted with $\epsilon$ in the following section.

\section{Strength of the signal and deformation of the limit cycle}

Recall the solution we found for the Bloch vector of the steady state of the synchronized two-level system in the rotating frame, Eq. \eqref{stationarySolution}. For small $\epsilon$ we can expand each of the Bloch vector components in powers of $\epsilon$

%
%

\begin{dgroup}
\begin{dmath}\label{XApprox}
m_x \approx A\cdot\epsilon\left[1 - K\epsilon^2\right]
\end{dmath}

\begin{dmath}\label{YApprox}
m_y \approx  B\cdot\epsilon\left[1 - K\epsilon^2\right]
\end{dmath}

\begin{dmath}\label{ZApprox}
m_z \approx  C\cdot\left[1 - K\epsilon^2\right]
\end{dmath}
\end{dgroup}
where
\begin{align}
    A &= \frac{4 (\Gamma _g - \Gamma _d)}{\left(\Gamma_d + \Gamma_g\right)^2 + 16\Delta^2}\\
    B &= \frac{16 \Delta(\Gamma _g - \Gamma _d)}{(\Gamma_d + \Gamma _g)[(\Gamma _d + \Gamma _g)^2 + 16\Delta ^2]} \\
    C &= \frac{\Gamma_d - \Gamma_g}{\Gamma_d + \Gamma_g}\\
    K &= \frac{8}{(\Gamma _d + \Gamma _g)^2 + 16\Delta ^2}
\end{align}
Note that the first non-constant term of \eqref{ZApprox} is quadratic in $\epsilon$ while both \eqref{XApprox} and \eqref{YApprox} are linear in $\epsilon$. Thus, we expect that for small signal strengths, the Bloch vector components that change with $\epsilon$ are $m_x$ and $m_y$, showing that the system phase-locks to the external signal. $m_z$  will remain approximately constant, and it will have the value it had without signal.
 In general, we could say that when the $z$-component of the Bloch vector moves far from its original non-signal value, the limit cycle is perturbed because the signal is too strong, and therefore we are not talking about synchronization anymore. Thus, the parameter that tells us if our system is forced or not is $K\epsilon^2$, as some kind of \textit{deformation parameter}.

Let us consider the case of Fig. \ref{QFd}. Here,
$K\epsilon^2 \approx 0.12$, and therefore $m_z \approx 0.88\cdot C$. We observe that the value of $m_z$ is close to $C$, but it is not that close. In this regime, the limit cycle is slightly deformed, although we could consider it is still a valid picture for synchronization. 
In Figures \ref{ATa}, \ref{ATb}, \ref{ATc}, the maximum value of the strength is $\epsilon = 1$, which corresponds to $K\epsilon^2 = 0.07$. The boundary between synchronization and forced oscillation is not sharply defined, but it seems reasonable to say  that we are in the synchronization regime for the parameters values used in these figures.

\section{General three-level system model and appearance of limit cycle}

In this section we want to contrast the TLS-model that we have studied with the three-level model of \cite{paper, paper2, paper3}, and see how to modify this so that it will have a limit cycle in the same sense as we have described (see Fig. \ref{2DBS}). 
The key difference between these two systems is the nature of the steady state in the absence of signal. In order to explain the fact that the two-level system can be synchronized we have made use of the interpretation of the mixed state as a probabilistic ensemble of pure states, each of them describing a different and valid limit cycle. In this way, this system is analogous to a classical system that synchronizes in the presence of noise.

In the  three-level model studied in  \cite{paper},  the steady state is the pure state $\ket{0}$ which is an eigenstate of the Hamiltonian.   It has a Husimi Q distribution that is equally distributed over all $\phi$, which means that it is a quantum superposition of a set of spin coherent states, each of which are moving along the limit cycle. The crucial difference between our approach and that of Ref.~\cite{paper} is therefore that  we allow statistical
mixtures instead of quantum superpositions in the resolution of the
stationary state in terms of pure states evolving on a limit
cycle. This allows a unified understanding of the synchronization behaviour of two- and three-level systems.  

In this framework, the stationary state $\ket{0}$ of the Lindblad equation for the three-level model in Ref.~\cite{paper} is not interpreted as a limit cycle, being one single eigenstate of the Hamiltonian. 
Only when adding a signal, the stationary state is pushed away from this eigenstate and we obtain a cycle. However, in this case it seems natural to tell that there is a substantial deformation of the stationary state, as the limit cycle changes from a single point to a circle. 

We can consider a more general three-level system which will show a limit cycle in the same sense as our TLS-model. If $\ket{m_z}$ are the eigenstates of the spin 1 operator $S_z$, the model of Ref.~\cite{paper} included transitions only from  $\ket{1}$ or $\ket{-1}$ to $\ket{0}$. In the extended model we allow transitions from $\ket{0}$ to $\ket{1}$ or $\ket{-1}$  in addtion.

The modified Lindblad equation in the absence of an external signal would be similar to that of the TLS, although we need to work with a generalized eight-dimesional Bloch sphere if we want to study the motion of the Bloch vector of the system \cite{goyal}:

\begin{equation}
\frac{d\hat{\rho}}{dt} = \alpha \mathcal{D}[\hat{S}_+]\hat{\rho} + \beta\mathcal{D}[\hat{S}_-]\hat{\rho}
\end{equation}
where $\alpha$ and $\beta$ are transition rates. The form of the density matrix for a three-level system is

\begin{dmath}[style = {\small}]
\hat{\rho} =
\begin{pmatrix}
\frac{1}{\sqrt{3}} + m_3 + \frac{1}{\sqrt{3}}m_8 & m_1-im_2 & m_4 + im_5 \\
m_1 + im_2 & \frac{1}{\sqrt{3}} - m_3 + \frac{1}{\sqrt{3}}m_8 & m_6 - im_7\\
m_4 + im_5 & m_6 + im_7 & \frac{1}{\sqrt{3}} - \frac{2}{\sqrt{3}}m_8
\end{pmatrix}
\end{dmath}
The stationary solution is

\begin{equation}
m_1 = m_2 = m_4 = m_5 = m_6 = m_7 = 0
\end{equation}

\begin{dmath}[style = {\small}]
m_3 = \frac{-12 \alpha^2-6 \alpha \beta }{\sqrt{3} (2 \alpha+4 \beta) (2 \alpha+\beta)}
\end{dmath}

\begin{dmath}[style = {\small}]
m_8 = \frac{1}{2} + \frac{ \beta(\alpha - 4 \beta)}{2 \alpha (2 \alpha+\beta)}
\end{dmath}

The Bloch vector that corresponds to the $\ket{0}$ state is $m_3 = -\frac{\sqrt{3}}{2}$, $m_8 = \frac{1}{2}$, and the rest of the components are equal to zero. We see that, in general, the steady state will differ from this one and  is instead a mixed state. Thus, in this case we have a mixture of pure states on a circle similar to the one in  Fig. \ref{2DBS}, which we can see as the analogue of the limit cycle in a classical system. Notice that not all values of $\alpha$ and $\beta$ are allowed since we need to keep our states inside an eight-dimensional sphere of radius one \cite{goyal}. 

\section{Conclusions}

We have shown that, inside the classical synchronization framework, it is possible to understand that a two-level system provides a valid limit cycle if we interpret mixed states as a probability mixture of pure states with a limit cycle associated to each of them. Explaining the appearance of this cycle was the missing point in previous works \cite{ExtraPaper3, ExtraPaper4}, being essential for making an analogy with the classical counterpart of synchronization. Equations for the two-level system can be analytically solved and the motion of the Bloch vector in the presence of an external signal can be obtained. Therefore, synchronization can be achieved, but without full phase-locking (as is the case for every quantum system that synchronizes, due to quantum noise). Also, the Husimi Q representation is a powerful tool for characterizing the synchronization regimes and strength of phase-locking.
We also studied the evolution and distortion of the limit cycle with the strength of the signal, since it is fundamental that the signal is weak not to move away from the synchronization regime, as defined classically \cite{book}.

With these results for the TLS we can compare it to the three-level system treated in \cite{paper, paper2, paper3}. We observe that the response of the system to an external signal is essentially the same for both systems, showing that it is reasonable to consider a qubit to be synchronized in a similar way as a three-level system. Our interpretation of the stationary state as a statistical mixture of pure states following the limit cycle enables a clear interpretation of this behaviour. In fact, a more general model for a three-level system than the one used in these previous works is needed if one attempts to understand synchronization starting from a similar limit cycle, since a mixed stationary state is required for following the same argument.

Being able to synchronize such small systems is of great interest because the qubit is the basic unit of quantum computation. Quantum information theory has been in constant development these last years, and hence learning how these qubits can synchronize can be useful in the quantum computing field. We beleive that clarifying the mechanisms by which limit cycles, characteristic of self-sustained oscilltors, arise in quantum systems will help further studies about quantum synchronization.

\begin{acknowledgments}
We thank Jebarathinam Chellasamy for bringing Refs. \cite{TPaper1,TPaper2} to our attention. 
\end{acknowledgments}


\end{document}